\documentclass[12pt]{article} 
\usepackage[tablesfirst,notablist,nomarkers]{endfloat} 

\usepackage{ol2}
\usepackage[draft]{hyperref}
\usepackage{amsmath}

\begin{document}

\title{{Delayed four- and six-wave mixing in a \\ coherently prepared atomic ensemble}}

\author{D.~Felinto, D.~Moretti${^\dagger}$, R.~A.~de Oliveira, and J.~W.~R.~ Tabosa${^*}$}

\affiliation{Departamento de F\'{\i}sica, Universidade Federal de
Pernambuco,
\\ 50670-901 Recife, PE - Brazil \\
$^*$Corresponding author: tabosa@df.ufpe.br}

\begin{abstract}We report on the simultaneous observation, by delayed Bragg diffraction, of four- and six-wave mixing
processes in a coherently prepared atomic ensemble consisting of
cold cesium atoms. For each diffracted order, we observe different
temporal pulse shapes and dependencies with the intensities of the
exciting fields, evidencing the different mechanisms involved in
each process. The various observations are well described by a
simplified analytical theory, which considers the atomic system as
an ensemble of three-level atoms in $\Lambda$ configuration.
\end{abstract}

\ocis{270.1670, 190.4223, 020.1670}

\maketitle
The transfer of light coherence to atomic quantum-state coherence
is at the heart of many intriguing effects associated with
light-matter interactions and has played a fundamental role for the
enormous developments achieved in this field \cite{Lukin03}.
Coherently prepared atomic systems giving rise to electromagnetic
induced transparency (EIT), coherent population trapping, slow and
stopped light, have attracted much attention in recent years
\cite{Arimondo1996,Lukin00,Fleischhauer05}, specially owing to the
possibility of storing and manipulating classical and quantum
information in long-lived atomic coherences
\cite{Lukin03}. Nonlinear cw four-wave
mixing (FWM) processes in EIT media have been extensively
investigated \cite{Kumar1997, Tabosa02} and, in particular, they have been applied for the
generation, in atomic ensembles, of both narrow-band photon pairs \cite{Harris06} and pairs of intense light beams showing a
high degree of intensity squeezing \cite{Lett07}. On the other
hand, multi-wave mixing processes have also been investigated
using spatially resolved techniques\cite{Cardoso02,Kang04} and, more
recently, the interference between cw FWM and six wave mixing
(SWM) signals was
demonstrated \cite{Xiao09}.

In this Letter, we report the observation of higher order
nonlinear wave mixing processes in a coherently prepared cold
atomic ensemble by means of the simultaneous emission of two
pulses generated by delayed FWM and SWM processes. The delayed FWM pulse generation is presently a well known phenomenon, which
has been explored particularly for applications in all-optical routing~\cite{Ham08} and in quantum
information~\cite{Chou2007}. The delayed SWM
pulse generation, on the other hand, was just recently predicted
in~\cite{Moretti10} and here we provide its first observation,
highlighting its difference with the corresponding FWM process by
detecting both signals simultaneously.
Different pulse shapes and dependencies on the incident field
intensities are observed. All these features are predicted by a
simplified theoretical model which allowed us to characterize the
two pulses as associated with distinct nonparametric, nonlinear
processes. The possibility of observing higher order nonlinear
interaction between light and coherent atomic systems could play
an important role for applications in quantum information, since
they present quite different photon correlations. One goal, for
example, is to generate quantum-correlated pulse pairs from such
coherently prepared atomic ensembles~\cite{Moretti10}. The delayed
FWM pulse pairs should exhibit correlations only in the saturated
regime~\cite{Moretti10}, the pulses being independent from each
other at low powers. On the other hand, a pair of delayed SMW
pulses should exhibit stronger correlations for any readout power,
since they rely on all fields to be generated.

The theoretical analysis of the problem was developed in
\cite{Moretti10} and here we just review its main steps. We
consider three-level atoms consisting of two degenerate
Zeeman ground states and one excited Zeeman state interacting with
two writing beams $W$ and $W^{\prime}$ with a small angle between
them and with opposite circular polarizations (Fig. 1). These two beams act for a long time so as to
induce a stationary Zeeman ground state coherence grating before
they are turned off. After the storage time $t_s$, the grating is
read out by two reading beams $R$ and $R^{\prime}$, which also have
opposite circular polarizations (Fig. 1). In our model, the stored coherence grating decays
with a rate $\gamma$ assumed to be much smaller than the excited
state decay rate $\Gamma$.

As a result of the action of these four fields, optical coherences
are excited in both transitions, resulting in the emission of two
fields with opposite polarizations and propagating in the
direction opposite to $W^{\prime}$. In \cite{Moretti10} we show
that these optical coherences can be written as a sum of three
terms, corresponding to the processes of stimulated emission,
delayed FWM, and delayed SWM, respectively. Not all these processes, however, are phase
matched and result in propagating optical fields. For the
experimental situation considered here, note that the two reading
beams are both counterpropagating with respect to $W$. This
geometry for the readout fields is quite different from the one
explored in \cite{Moretti10}, and results in the phase matching of
one of the delayed SWM terms present in the locally excited
optical coherences, together with one of the delayed FWM terms
explored in \cite{Moretti10}. Explicitly, from
\cite{Moretti10} we obtain that the terms of the local optical
coherences contributing to these phase-matched, delayed FWM and
SWM signals are given respectively by
\begin{subequations}
\begin{align}
\sigma_{M_{F}-1,M_{F}}(t) \propto &\; \frac{|\Omega_R|e^{-\gamma t_s}}{I_t \Gamma} \left[ f_r(t)I_R + g_r(t) I_{R^{\prime}} \right] \nonumber \\
                  & \;\;\;\; \times e^{-i(\vec{k}_R+\vec{k}_W-\vec{k}_{W^{\prime}})\cdot \vec{r}} \;, \\
\sigma_{M_{F}+1,M_{F}}(t) \propto &\; \frac{I_R |\Omega_{R^{\prime}}| e^{-\gamma t_s}}{I_t \Gamma} \left[ f_r(t) - g_r(t) \right] \nonumber \\
                  & \;\;\;\; \times e^{-i(2\vec{k}_R-\vec{k}_{R^{\prime}}+\vec{k}_W-\vec{k}_{W^{\prime}})\cdot \vec{r}}\;,
\end{align}
\label{sigmas}
\end{subequations}
where $\vec{k}_X$ (with $X=$ $R$, $R^{\prime}$, $W$, or
$W^{\prime}$) are wavevectors associated with the reading and
writing beams, considered as plane waves all having the same
frequency, resonant with the atomic transition. These expressions
are valid for $\gamma$ much smaller than the Rabi frequencies
$\Omega_X$ \cite{Moretti10}. Since $\gamma << \Gamma$ for our
experimental conditions, they describe both low-intensity and
saturated regimes of these processes. For simplicity, we omitted
the dependence with $\Omega_W$,$\Omega_{W^{\prime}}$ from these
expressions. An analysis of their low-intensity limit reveals the
usual dependences of $|\Omega_W||\Omega_{W^{\prime}}||\Omega_R|$
and
$|\Omega_W||\Omega_{W^{\prime}}||\Omega_R|^2|\Omega_{R^{\prime}}|$
for the FWM and SWM processes, respectively. In
Eqs.~\eqref{sigmas}, $I_R$ ($I_{R^{\prime}}$) corresponds to the
intensity of beam $R$ ($R^{\prime}$), and
$I_{t}=I_{R}+I_{R^{\prime}}$. Functions $f_{r}(t)$ and $g_{r}(t)$
are defined in \cite{Moretti10}. They depend only on the
parameters $I_{t}$ and $\Gamma$ and determine the temporal pulse
shapes associated with each term of the optical coherences.

\begin{figure}[htb]
  \centerline{\includegraphics[width=5.2cm, angle = -90]{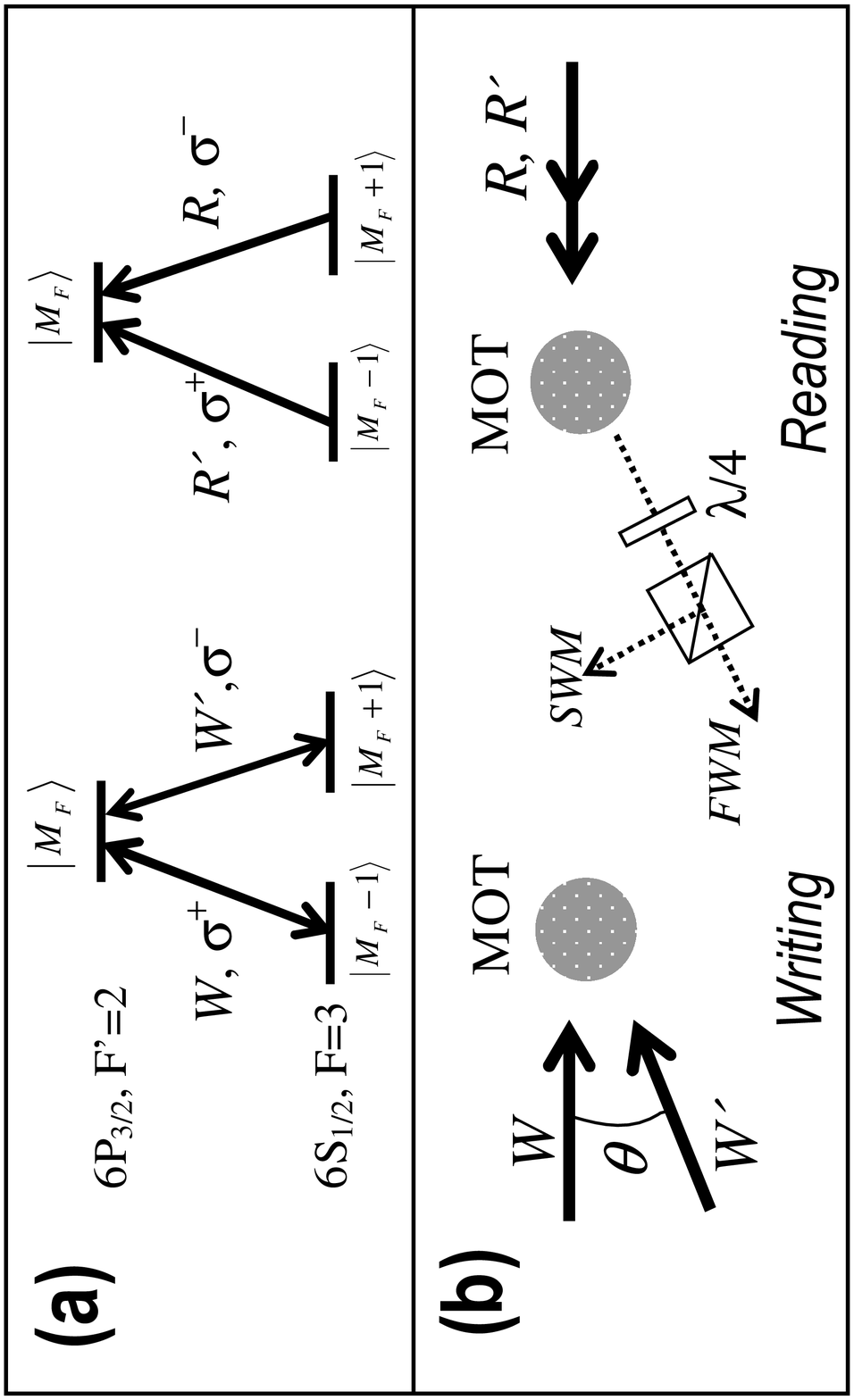}}
  \vspace{-0.0cm}
  \caption{ (a) Zeeman three-level system considered in the theoretical model,
indicating the coupling of each state with the respective writing
and reading fields. (b) Propagation directions of writing ($W$,$W^{\prime}$), reading
($R$,$R^{\prime}$), and diffracted ($FWM$,$SWM$) beams.}
  \label{fig:Fig1}
\end{figure}

The delayed FWM signal in the above equations can be interpreted
as originating from Bragg diffraction of beam $R$ from the ground
state coherence grating created by the writing beams. On the other
hand, the SWM signal is a higher order process originated from
Bragg diffraction of beam $R$ from population gratings created in
the ground and excited states of the corresponding transition, by
the simultaneous action of both $R$ and $R^{\prime}$ over the
original coherence grating induced by the writing beams $W$ and
$W^{\prime}$. These population gratings responsible by the SWM
signal require then all four incident fields in Fig.~1(a) to be
created, while the coherence grating responsible by the FWM signal
requires only fields $W$, $W^{\prime}$.

Since we consider a system composed of
Zeeman sublevels, each transition corresponds to a well defined
selection rule. Therefore the beams generated in each transition
have orthogonal circular polarizations, an essencial aspect for
separating each signal experimentally. Neglecting propagation
effects, the electric field and the intensity in each mode can be
promptly calculated \cite{Moretti08}. Experimentally we focus
attention on the pulse shapes associated with the FWM and SWM
processes (proportional to the squared modulus of Eqs. (1a) and
(1b), respectively) and on the corresponding retrieved energies. The expressions for
the retrieved energies in each pulse, that will be used to compare with our experimental results, can be readily obtained
\cite{Moretti08,Moretti10} from Eqs. (1a) and (1b):
\begin{subequations}
\begin{align}
U_{FWM} & \;\propto \frac{I_R}{I_t2} \int_0^{\infty} \left| f_r(t)I_R + g_r(t) I_{R^{\prime}}\right|^2 dt \;, \\
U_{SWM} &\propto \frac{I_{R^{\prime}} I_R2}{I_t2} \int_0^{\infty}
|f_r(t)-g_r(t)|^2 dt \;.
\end{align}
\label{Us}
\end{subequations}

The experiment is performed using cold cesium atoms
provided by a magneto-optical trap (MOT). The atoms are initially
pumped into the $6S_{1/2}(F=3)$ ground state using the method
described in \cite{Moretti08}. The MOT temperature was estimated
in the range of mK and the atomic density in the lower ground
state is $n\approx 10^{10}$ cm$^{-3}$. The MOT quadrupole magnetic
field is switched off while three pairs of Helmholtz coils are
used in order to compensate for stray magnetic fields. For the
writing and reading beams we employ an external cavity diode laser
locked on the cesium closed hyperfine transition
$6S_{1/2}(F=3)\rightarrow 6P_{3/2}(F^{\prime }=2)$, with the time sequence determined by acousto-optical modulators. The opposite
circular polarization states for the pairs of writing and reading
beams are obtained by using polarizing beam splitters (PBS) and quarter
waveplates. The grating storage time $t_s$ is
fixed and equal to $1~\mu s$. The manipulation and storage of such
gratings in our system, as a function of $t_s$, were investigated
in more detail in~\cite{Moretti08,Moretti2010b}. The retrieved
signals associated with the delayed FWM and SWM can be separated
with a quarter waveplate followed by a PBS as
shown in Fig. 1(b).

In Fig. 2(a) we show the temporal shape of the retrieved pulses
associated with each wave mixing process. The corresponding powers
for the writing ($W$, $W^{\prime}$) and reading ($R$,
$R^{\prime}$) beams are approximately equal to (20 mW/cm$^2$, 2
mW/cm$^2$) and (10 mW/cm$^{2}$, 10 mW/cm$^{2}$), respectively.
Clearly, the different observed pulse shapes strongly suggest the
two signals originate from different mechanisms. In Fig. 2(b) we
plot the theoretical pulse shapes associated with each retrieved
signal. In order to compare directly with the experimental
results, we convoluted the pulse shapes provided by Eqs.~(1a)
and~(1b) with the time response of our detectors (PDA36A from
Thorlabs), described by a single response-time parameter
$\tau$~\cite{Butterworth1968}. From such convolution we estimate
$\tau = 0.2\;\mu$s, consistent with the detector's bandwidth.

\begin{figure}[htb]
  \centerline{\includegraphics[width=5.5cm, angle=-90]{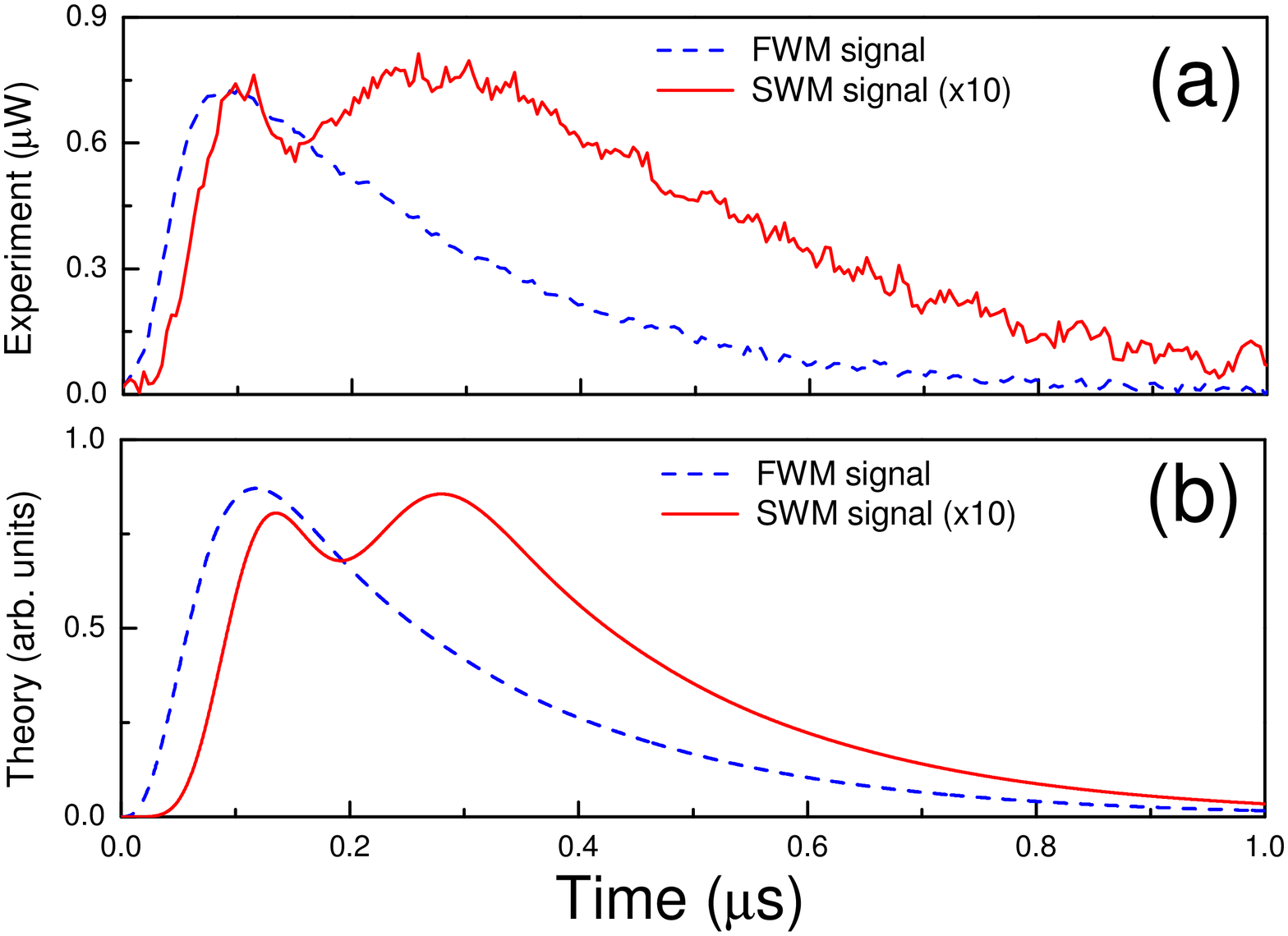}}
  \vspace{-0.2cm}
  \caption{(a) Measured temporal pulse shapes associated with the delayed FWM and SWM processes.
  (b) Calculated pulse shapes for $I_t = 2.8\,I_s$ and $\tau = 0.2\;\mu$s, with $I_s$ the saturation intensity of the transitions.}
    \label{fig:Fig2}
\end{figure}

\begin{figure}[htb]
  \centerline{\includegraphics[width=3.8cm, angle=-90]{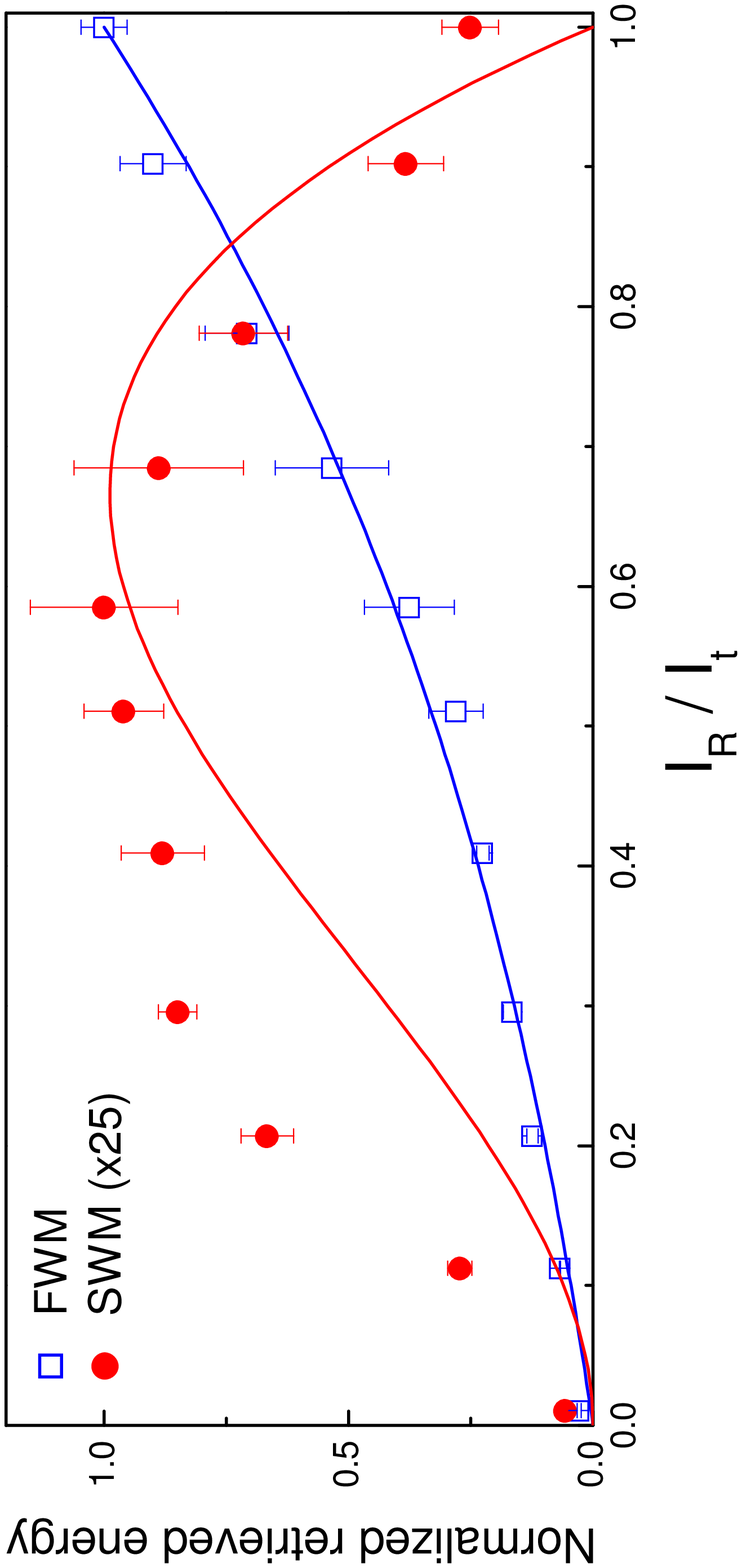}}
  \vspace{-0.2cm}
  \caption{Retrieved energy associated with the FWM (open squares) and the SWM (filled circles) processes.
  The solid lines are the theoretical curves from Eqs. (\ref{Us}) with $I_t = 2.2\,I_s$. Both theoretical and experimental curves were normalized by the maximum retrieved energy $U_{max}$, corresponding to the FWM pulse at $I_R = I_t$. For the experiment, we had $U_{max} = 1.3$~pJ.}
  \label{fig:Fig3}
\end{figure}

In another series of experiments, we fixed the total intensity
$I_t$ associated with the reading beams and changed the relative
values of $I_{R}$ and  $I_{R^{\prime}}$ (so that $I_{R^{\prime}} =
I_t - I_R$). In Fig. 3 we show the retrieved pulse energy
associated with the FWM and SWM processes as a function of the
intensity of field $R$ normalized by $I_t$. As expected, while the
extracted energy associated with the FWM process monotonically
increases with the reading beam intensity $I_{R}$, the similar
quantity associated with the SWM process vanishes at the extreme
values of the intensities corresponding to ($I_{R}=0$,
$I_{R}^{\prime}=I_t$) and ($I_{R}=I_{t}$, $I_{R^{\prime}}=0$),
respectively. This behavior is consistent with the predictions of
Eq. 2(b), which require all fields to be present for the SWM
signal to be observed. The theoretical curves, solid lines,
describe quite well these observations, together with the relative
values of the retrieved energy for each process. Even though, they
fail to describe the observed symmetry of the curve related to
$U_{SWM}$. We attribute this discrepancy to the simplicity of our
model with respect to the real Zemann structure in the experiment.

In summary, we have demonstrated experimentally the observation of
delayed four- and six-wave mixing in a coherently prepared cold
cesium ensemble. The measured results are in qualitative agreement
with the predictions of a simple theoretical model for an ensemble
of three-level atoms. Our results clearly demonstrate the
possibility for distributing previously stored light information
in different optical modes and through different nonlinear
processes, which might be of considerable interest for the growing
field of quantum information.

This work was supported by the Brazilian agencies CNPq and FACEPE
and the National Institute of Science and Technology for Quantum
Information.

$\dagger$ $Present ~address$: Departamento de F\'{\i}sica,
Universidade Federal de Campina Grande, PB, Brazil.

\end{document}